\documentclass[journal,twoside]{IEEEtran}

\usepackage{float}
\usepackage{stackengine}
\usepackage{setspace}
\usepackage{amssymb,amsfonts,latexsym}
\usepackage{graphicx}
\usepackage{subfigure}
\usepackage{psfrag}
\usepackage{algorithmicx}
\usepackage{blindtext}
\usepackage{comment}
\usepackage{cite}
\usepackage{url}
\usepackage{times}
\usepackage{bbm}
\usepackage{epsfig}
\usepackage{multirow}
\usepackage{longtable}
\usepackage{amsfonts}
\usepackage{amssymb}%
\usepackage{color}
\usepackage{mathrsfs}
\usepackage{bm}
\usepackage{booktabs}
\usepackage{threeparttable}
\usepackage[cmex10]{amsmath}
\usepackage{amsmath}
\usepackage[english]{babel}
\usepackage{mathbbold}
\usepackage{mathtools, nccmath}
\usepackage[linesnumbered,ruled,noend]{algorithm2e}
\makeatother
\usepackage[table]{xcolor}
\usepackage{makecell}
\usepackage{stackengine}

\usepackage{stackengine}

\usepackage{stfloats}

\ifCLASSINFOpdf

\else

\fi

\begin{document}

		\title{Cost-Effective Two-Stage Network Slicing for Edge-Cloud Orchestrated Vehicular Networks}
\author{\IEEEauthorblockN{Wen Wu$^\ddagger$, %Conghao Zhou$^\star$, 
		Kaige Qu$^\star$, Peng Yang$^*$, % Qiang Ye$^\diamond$, 
		Ning  Zhang$^\dagger$,  Xuemin (Sherman) Shen$^\star$, and Weihua Zhuang$^\star$}
	\\\IEEEauthorblockA{
	Frontier Research Center, Peng Cheng Laboratory, Shenzhen, China$^\ddagger$\\
	Department of Electrical and Computer Engineering, University of Waterloo, Waterloo, Canada$^\star$\\
	School of Electronic Information and Communications, Huazhong University of Science and Technology, China$^*$\\
%	Department of Computer Science, Memorial University of Newfoundland, St. John’s,  Canada$^\diamond$\\
			Department of Electrical and Computer Engineering, University of Windsor, Windsor, Canada$^\dagger$\\
		Email: wuw02@pcl.ac.cn$^\ddagger$, \{%c89zhou, 
		k2qu, sshen, wzhuang\}@uwaterloo.ca$^\star$}, \\yangpeng@hust.edu.cn$^*$, and ning.zhang@uwindsor.ca$^\dagger$  }
\maketitle
\begin{abstract}

%Network slicing can support diversified Internet of vehicles services by constructing multiple logically-isolated virtual networks, namely slices, on top of a common physical network. 
%In this paper, we study network slicing problem for edge-cloud orchestrated vehicular networks, in which the edge and cloud servers are orchestrated to process computation tasks for reducing network slicing cost while satisfying the quality of service requirements. We propose a two-stage network slicing scheme, which consists of 1) network planning stage in the large timescale to perform slice deployment, edge resource provisioning, and cloud resource provisioning, and 2) network operation stage in the small timescale to perform resource allocation and task dispatching. Particularly, we formulate the network slicing problem as two-timescale stochastic optimization problem to minimize the network slicing cost. Since the problem is NP-hard, we develop a Two timescAle netWork Slicing (TAWS) algorithm by collaborating reinforcement learning (RL) and optimization methods, which can jointly make network planning and operation decisions. Specifically, by leveraging the timescale separation property of decisions, we decouple the problem into a large-timescale network planning subproblem and a small-timescale network operation subproblem. The former is solved by an RL method, and the latter is solved by an optimization method. Simulation results based on real-world vehicle traffic traces show that the TAWS can effectively reduce the network slicing cost as compared to benchmark schemes.
In this paper, we study a network slicing problem for edge-cloud orchestrated vehicular networks, in which the edge and cloud servers are orchestrated to process computation tasks for reducing network slicing cost while satisfying the quality of service requirements. We propose a two-stage network slicing framework, which consists of 1) \emph{network planning} stage in a large timescale to perform slice deployment, edge resource provisioning, and cloud resource provisioning, and 2) \emph{network operation} stage in a small timescale to perform resource allocation and task dispatching.  Particularly, we formulate the network slicing problem as  a two-timescale stochastic optimization problem to minimize the network slicing cost. Since the problem is NP-hard due to coupled network planning and network operation stages, we develop a \underline{T}wo timesc\underline{A}le net\underline{W}ork \underline{S}licing (TAWS) algorithm by collaboratively integrating reinforcement learning (RL) and optimization methods, which can jointly make network planning and operation decisions. Specifically, by leveraging the timescale separation property of decisions, we decouple the problem into a large-timescale network planning subproblem and a small-timescale network operation subproblem. The former is solved by an RL method, and the latter is solved by an optimization method.  Simulation results based on real-world vehicle traffic traces show that the TAWS can effectively reduce the network slicing cost as compared to the benchmark scheme.

\end{abstract}

%------------------------------------------------------------------------------------------------------------------------------------------------
\section{Introduction}

%For example, a safety message dissemination service requires high reliability (e.g., 99.999\%), a mobile video streaming service requires high throughput (e.g., 5\;Gbps), and an environment perception service requires not only low latency (e.g., 100\;ms) but also high detection accuracy (e.g., 99.9\%). 
To make autonomous driving from a mere vision to reality, future vehicular networks are required to support various Internet of vehicles (IoV) services, such as object detection, in-vehicle infotainment, and safety message dissemination~\cite{campolo20175g}. Those IoV services have diversified quality of service (QoS) requirements in terms of delay, throughput, reliability, etc.  Emerging network slicing is deemed as a \emph{de-facto} solution to support diversified IoV services in vehicular networks. Its basic idea is to construct multiple isolated logical sub-networks (i.e., slices) for different services on top of the physical network, thereby facilitating flexible,
agile, and cost-effective service provisioning. Starting from the fifth-generation (5G) era,  standardization efforts from the 3rd generation partnership project (3GPP) body, e.g., Releases~15-17~\cite{kaloxylos2018survey, 3GPP2017, 3GPP2017_2}, and  proof-of-concept systems, e.g., Orion~\cite{foukas2017orion}, have fuelled the maturity of network slicing. In the coming 6G era, advanced network slicing techniques are expected to play an increasingly important role~\cite{you2021towards, shen2021holistic, wu2021ai}.   %(1) \emph{network planning}, where network resources are reserved to slices in a large timescale (e.g., from several minutes to several hours) for service provisioning, and (2) \emph{network operation}, where the reserved resources of a slice are further allocated to its subscribed end users in a small timescale (e.g., from several milliseconds to several seconds).
 
% The optimal RAN slicing policy requires jointly determining planning and operation decisions to optimize RAN slicing performance while satisfying QoS requirements. 

In the literature, significant research efforts have been devoted to network slicing. Ye \emph{et al.} investigated a radio spectrum resource slicing problem, in which radio spectrum is sliced between macro base stations (MBSs) and small BSs (SBSs)~\cite{ye2018dynamic}. To achieve efficient resource allocation, a deep learning-based algorithm was proposed to jointly allocate radio spectrum and transmit power in a slicing-based network~\cite{mei2021intelligent}. The previous work in ~\cite{wu2020dynamic} considered the resource provisioning problem and proposed a constrained learning algorithm to solve it. However, this work differs from the existing works in several important aspects. Firstly, the existing works focus on utilizing resources on the network edge, low-cost cloud resources are yet to be considered. As a remedy, a certain amount of computation tasks processed at the congested BSs can be dispatched to the remote cloud, i.e., \emph{task dispatching}, such that  system cost can be reduced. Secondly, network slicing includes two stages: 1) \emph{network planning} stage to provision network resources for slices in the large timescale, and 2) \emph{network operation} stage to allocate the reserved resources  to end users in the small timescale~\cite{3GPP2017, shen2020ai}. The existing works mainly decouple network slicing into two independent stages, while the interaction between them is seldom considered. Hence, designing a cost-effective network slicing scheme should take cloud resources and such interaction relationship into consideration.

%In addition, operating slice at a BS requires virtualizing network resources, which incurs a significant network slicing cost. Due to overlapped coverage among BSs, appropriately deploying slices at BSs, i.e., \emph{slice deployment}, can reduce network slicing cost while guaranteeing  slice coverage. Efficient network slicing schemes should incorporate both task dispatching ancloud slice deployment mechanisms.

%In addition, the existing works focus on optimizing the performance in either network planning stage or network operation stage. The optimal RAN slicing policy requires jointly optimizing planning and operation decisions to optimize system performance while satisfying QoS requirements. 

%In this paper, we introduce an edge-cloud orchestrated vehicular network. Taking the task dispatching mechanism into consideration, we propose a two-stage network slicing framework to support IoV services with diversified QoS requirements. 

Optimizing network slicing performance in dynamic vehicular networks faces the following \emph{challenges}. Firstly, network planning and operation decisions are \emph{nested}. Large-timescale network planning decisions (e.g., resource reservation), will condition small-timescale network operation decisions (e.g., resource allocation). Meanwhile, the performance achieved in the network operation stage will also affect the decision-making in the network planning stage, which is difficult to be solved by conventional optimization methods. Secondly, since vehicle traffic density varies temporal-spatially, network planning decisions need to be made to optimize long-term performance in the slice lifecycle while accommodating such network \emph{dynamics}. Deep reinforcement learning (RL) is considered as a plausible solution for long-term stochastic optimization. %Therefore, network planning and operation decisions should be jointly determined to achieve optimal performance in the presence of network dynamics.

%which can not optimize the overall performance. Since , we present a two-stage network slicing framework to automatic optimize the overall performaradionce. 

%1. incorporate cloud computing 
%2. two-stage: resource  

%The RL methods can accommodate time-varying environment to optimize long-term system performance.  

%When a vehicle is driving on the road, computation-intensive tasks are dispatched to the roadside network edge or remote cloud for prompt processing.

In this paper, we \emph{first} propose a cost-effective two-stage network slicing framework for edge-cloud orchestrated vehicular networks, by considering nested network planning and operation stages and effectively leveraging cloud resources. We then apply a network slicing cost model that accounts for slice deployment, resource provision, slice configuration adjustment, and QoS satisfaction. Based on the model, we formulate the network slicing problem as a two-timescale stochastic optimization problem to minimize the network slicing cost.  \emph{Second}, to solve the problem, we develop a learning-based algorithm, named \underline{T}wo timesc\underline{A}le net\underline{W}ork \underline{S}licing (TAWS).  The TAWS exploits the timescale separation structure of decision variables and decouples the problem into two subproblems in different timescales. Regarding the large-timescale network planning subproblem, an RL algorithm is designed to minimize network slicing cost via optimizing slice deployment, edge resource provisioning, and cloud resource provisioning. Regarding the small-timescale network operation subproblem, an optimization algorithm is designed to minimize average service delay via optimizing resource allocation and task dispatching. In addition, the achieved service delay in the network operation stage is incorporated into the reward of the RL-based network planning algorithm, thereby capturing the interaction between two stages and enabling \emph{closed-loop} network control. Simulation results on real-world vehicle traces demonstrate that the proposed algorithm outperforms the benchmark scheme in terms of reducing network slicing cost. 
%Our contributions in this paper are summarized as follows:

%problem formulation

%Considering network dynamics due to vehicle mobility and random task arrivals, we then formulate an optimization problem 

% considering both planning and operation stages. 

%We consider urban vehicular networks, in which vehicles moving on the road can offload their computation-intensive tasks to roadside base stations (BSs) equipped with mobile edge computing (MEC) servers for prompt processing. 

% Accordingly, we propose a two-timescale network slicing algorithm, named \underline{T}wo timesc\underline{A}le Net\underline{W}ork \underline{S}licing (TAWS). 

%\begin{itemize}
%	\item We propose a two timescale network slicing framework for vehicular networks;
%\item We present an edge-cloud orchestrated vehicular network architecture to facilitate load balance among roadside BSs;   
%\item We propose a two-stage network slicing framework and formulate a two-timescale stochastic optimization problem;
%\item We develop a learning-based algorithm, named TAWS, by collaboratively integrating RL  and optimization methods.

%	\item \textcolor{blue}{We may conduct simulations with imperfect traffic estimation};%We derive the optimal planning window size;
%\end{itemize}

%---------------------------
The remainder of this paper is organized as follows. The system model and problem formulation are presented in Sections~\ref{sec: system model} and~\ref{sec: problem_formulation}, respectively. Section~\ref{sec:solution} describes the proposed TAWS algorithm.  Simulation results are given in Section~\ref{sec: simulation results}, along with the conclusion in Section~\ref{sec: conclusion}.

%------------------------------------------------------------------------------------------------------------------------------------------------
\section{System Model}\label{sec: system model}
\subsection{Network Model}
\begin{figure}[t]
				\vspace{-0.2 cm}
	\centering
	\renewcommand{\figurename}{Fig.}
	\includegraphics[width=0.25\textwidth]{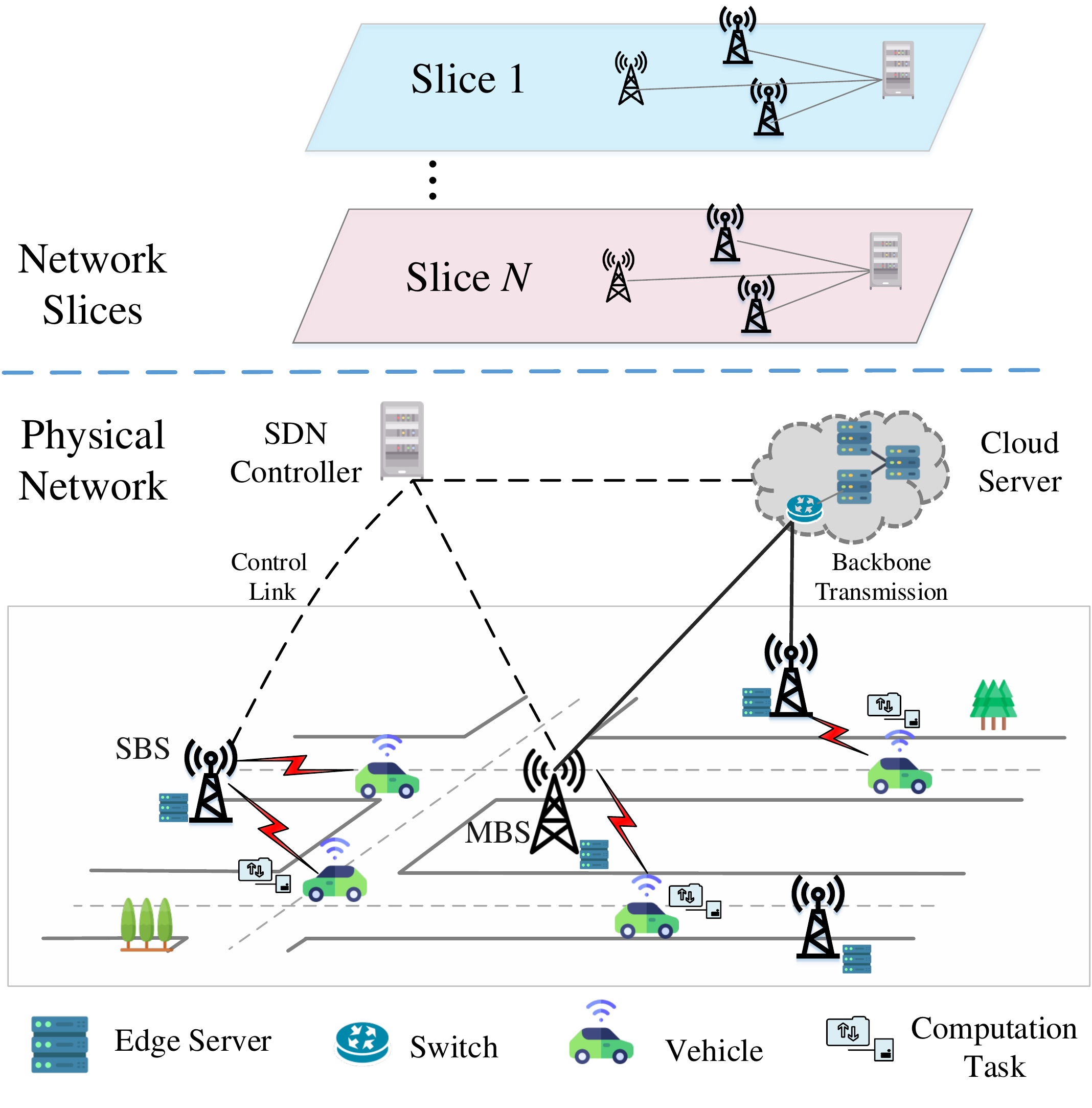}

	\caption{Network slicing for edge-cloud orchestrated vehicular networks.}
	\label{fig:system_model}
			\vspace{-0.7 cm}
\end{figure}

As shown in Fig.~\ref{fig:system_model}, the network slicing framework consists of several components.
%	\begin{itemize}
%		\item 

		{Physical network}: A two-tier cellular network is deployed for serving on-road vehicles. The set of BSs is denoted by $\mathcal{M}$, including the set of MBSs denoted by $\mathcal{M}_m$ and the set of SBSs denoted by $\mathcal{M}_s$, i.e., $\mathcal{M}= \mathcal{M}_m \cup \mathcal{M}_s$. Each BS has a  circular coverage and is equipped with an edge server. 
%		Regarding autonomous driving services, such as object detection and cooperative sensing, 
In the considered scenario, vehicles driving on the road generate computation tasks over time, which are offloaded to roadside BSs. Those tasks can be either processed at edge servers or dispatched to the remote cloud server via backbone networks. Once completed, computation results are sent back to vehicles.
		
%		\item 
		{Network slice}: Multiple network slices are constructed on top of the physical vehicular network. We consider $K$ delay-sensitive services with differentiated delay requirements, denoted by set $\mathcal{K}$. Let $\theta_k, \forall k \in \mathcal{K}$ denote the  tolerable delay of service $k$. For example, the tolerable delay of objective detection service is 100\;\emph{ms}~\cite{lin2018architectural}, whereas the tolerable delay of in-vehicle infotainment can be up to several hundreds of milliseconds.
		
%		\item 
		{Network controller}: A hierarchical network control architecture is adopted, including an upper-layer software defined networking (SDN) controller that connects to all BSs, and lower-layer local network controllers located at BSs. Those controllers are in charge of network information collection and making network slicing decisions.
%	\end{itemize}

%The slices' QoS requirements are differentiated in terms of service delay.%\footnote{Note that services can also be distinguished with different performance metrics.} 

  %For each slice, the local SDN controller located at the slice is in charge of the operation of the slice, e.g., user association and resource allocation.

%The centralized SDN controller connects to all the BSs, in charge of collecting stochastic network information, e.g., average vehicle traffic density, and making network planning decisions

% Local network controllers collect real-time network information, e.g., channel conditions, and determine network operation decisions of the BS.

\subsection{Two-Stage Network Slicing Framework}
We present a two-stage network slicing framework for the considered network. Firstly, a network planning stage operates in the large timescale (referred to as planning windows) to reserve resources at specific network nodes for the constructed slices. The  duration of each planning window is denoted by $T_p$.  At each planning window, the SDN controller collects the average vehicle traffic density information in the considered area, based on which planning decisions are made. Secondly, the network operation stage operates in the small timescale (referred to as operation slots) to dynamically allocate the reserved resources to vehicles according to real-time vehicles' service requests and network conditions. The duration of each operation slot is denoted by $T_o$. A planning window includes multiple operation slots,  i.e.,  $T_p/T_o \in \mathbb{Z}^+$. At each operation slot, the local network controller at each  BS collects real-time service requests and channel conditions  of  its  associated vehicles, based on which operation decisions are made.  Decision structures in two stages are detailed respectively as follows.

\subsubsection{Network Planning Decision Structure}  The planning window is indexed by $w\in\mathcal{W}= \{1, 2,..., W\}$, and planning decisions in planning window $w$ include the following components.

%	\begin{itemize}
%		\item 
		\emph{Slice deployment decision}, denoted by $\mathbf{o}^w\in \mathbb{R}^{M_s \times 1}$. Each element is a binary variable, i.e., 
		\begin{equation}\label{equ: slice deployment}
			o_{m}^w\in \{0, 1\},   m\in \mathcal{M}_s.
		\end{equation}	
		If  SBS $m$ is activated for slice deployment, we have $o_{m}^w=1$; otherwise,  $o_{ m}^w=0$. When service demands are low, deploying slices at a selective subset of  BSs can reduce network slicing cost as compared to deploying slices at all BSs while guaranteeing slices’ service level agreements (SLAs). This is because running network slicing requires resource virtualization, which incurs network operating costs. For service continuity consideration, we assume that MBSs that cover the entire area are always activated. Note that only when a BS is activated for slice deployment, edge resources at the BS can be provisioned.
		
%		The entire area $\mathcal{R}$ is divided into $J$ disjoint regions, denoted by $\mathcal{J}$, i.e., $\cup_{j=1}^{J} \mathcal{R}_j=\mathcal{R}$ where $\mathcal{R}_j$ denotes the $j$-th region. Since BSs are densely deployed, each region may be covered by a set of BSs, denoted by $\mathcal{B}_j  \subseteq \mathcal{B}, \forall j \in \mathcal{J}$. To guarantee seamless service coverage, at least one nearby BSs of region $j$ should be deployed with a slice, i.e., constraint 
%		\begin{equation}\label{equ:seamless_coverage_constraint}
%			\sum_{m \in \mathcal{B}_j }o_{m}^w>0, \forall k \in \mathcal{K},   j \in \mathcal{J}
%		\end{equation} 
%		should be satisfied;
		
%		\item 
		\emph{Edge resource provisioning decision}, including radio spectrum and computing resource provisioning at all BSs for all slices, denoted by $\mathbf{B}^w\in \mathbb{R}^{K \times M}$ and $\mathbf{C}^w\in \mathbb{R}^{K \times M}$, respectively. The corresponding elements
		\begin{equation}\label{equ:constraint_spectrum resource }
			\begin{split}
				\{b_{k,m}^w,c_{k,m}^w \} \in \mathbb{Z}^+,  \forall k\in \mathcal{K}, m\in \mathcal{M},
%				&c_{k,m}^w  \in \mathbb{Z}^+,  \forall k\in \mathcal{K}, m\in \mathcal{M}, w\in \mathcal{W}
			\end{split}
		\end{equation}
		represent the number of subcarriers and edge virtual machine (VM) instances provisioned for slice $k$ at BS $m$, where $\mathbb{Z}^+$ denotes the set of positive integers.\footnote{Memory resource is also allocated to the VM instance to enable task processing, which is matched to its allocated computing resource.} The bandwidth of a subcarrier is denoted by $\beta$, and the computing capability of  an edge VM is denoted by $F_e$.  Due to the limitation of edge resources, the following capacity constraints are imposed:
%		\begin{equation}\label{equ: spectrum capacity}
%			\begin{split}
%			 o_{m}^w \sum_{k\in \mathcal{K}}  b_{k,m}^w  &\leq B_m,  \forall m\in \mathcal{M},\\
%			  o_{m}^w	\sum_{k\in \mathcal{K}} c_{k,m}^w &\leq C_m,  \forall m\in \mathcal{M},
%			\end{split}
%		\end{equation}
		\begin{equation}\label{equ: spectrum capacity}
%	\begin{split}
		o_{m}^w \sum_{k\in \mathcal{K}}  b_{k,m}^w  \leq B_m, 
		o_{m}^w	\sum_{k\in \mathcal{K}} c_{k,m}^w \leq C_m,  \forall m\in \mathcal{M},
%	\end{split}
\end{equation}
		where $B_m$ and $C_m$ represent the total numbers of subcarriers and VM instances at BS $m$, respectively.
		
%			\item 
			\emph{Cloud resource provisioning decision}, denoted by $\mathbf{h}^w\in \mathbb{R}^{K \times 1}$. Each element
			\begin{equation}\label{equ: Cloud resource provision}
				h_k^w\in \mathbb{Z}^+, \forall k \in \mathcal{K}
			\end{equation}
		denotes the number of cloud VM instances reserved for slice~$k$. The computing capability of  a cloud VM is denoted by $F_c$. 
%		The decision vector is .

%	\end{itemize}
	
 \subsubsection{Network Operation Decision Structure} 
 Let $t \in  \mathcal{T}= \{1, 2,..., T\}$ denote the index of operation slots within a planning window.   At operation slot $t$, the following  decisions are determined for each slice $k$.
%\end{itemize}

%In operation slot $t$, the following network operation decisions should be determined.
%\begin{itemize}
%	\item \emph{User association decision}, denoted by $\mathbf{Z}^t \in \mathbb{R}^{{N}^t \times {M}}$. Each element 		
%	\begin{equation}\label{equ:user_association_binary_constriant}
%		z_{n, m}^t \in \{0, 1\}, \forall n\in \mathcal{N}^t, m\in \mathcal{M}^w 
%	\end{equation} is a binary variable to indicate whether vehicle $n$ is associated to BS $m$. Although a vehicle may be covered by multiple BSs, we consider a simple single-connectivity mode, i.e.,  each vehicle can only associate to one BS at an operation slot, yielding the following constraint:
%	\begin{equation}\label{equ:constriant_user_association-1}
%		\sum_{m\in \mathcal{M}}z_{n, m}^t=1, \forall n\in \mathcal{N}^t.
%	\end{equation}
%	
%	In addition, the user association decision should satisfy the physical connection constraint. Let  $\mathbf{E}^t \in \mathbb{R}^{{N}^t \times {M}}$ represent the connection status between vehicles and BSs at operation slot $t$. When user $n$ is covered by BS $m$, $e_{n,m}^t=1$; otherwise, $e_{n,m}^t=0$. As such, the following constraint should be satisfied:
%	\begin{equation}\label{equ:constriant_user_association-2}
%		z_{n, m}^t\leq e_{n,m}^t, \forall n\in \mathcal{N}^t, m \in\mathcal{M}^w .
%	\end{equation}
	%	which guarantees feasible user association decisions. 

	%The association decision is to dynamically adjust the queue backlog of each VM instance, thereby reduce system queuing delay. 

%	\item 
	\emph{Radio spectrum allocation decision}, denoted by $\mathbf{y}_k^t \in \mathbb{R}^{{N}^t \times 1}$. The reserved radio spectrum at each BS is allocated to active vehicles within BS's coverage for task offloading. Due to vehicle mobility, the number of  vehicles varies across time. Let $\mathcal{N}^t$ denote the set of active vehicles in operation slot $t$, and ${N}^t=|\mathcal{N}^t|$. For simplicity, each vehicle associates to the nearest BS. Let $\mathcal{N}^t_m$ denote the set of active vehicles associated to BS $m$ at operation slot $t$, and  $y_{k,n}^t \in \mathbb{R}^+$ represents the fraction of radio spectrum allocated to vehicle $n$.  The total amount of the allocated bandwidth should not exceed the reserved number of subcarriers at the corresponding BS, i.e.,
	\begin{equation}\label{equ:spectrum_constraint}
		\sum_{n\in \mathcal{N}^t_m }y_{k,n}^t \leq b_{k,m}^w,  \forall m \in\mathcal{M}^w.
	\end{equation}
	Here, $\mathcal{M}^w$ denotes the set of the activated BSs in window $w$.
	
%		\item 
			\emph{Task dispatching decision}, denoted by $\mathbf{x}_k^t\in \mathbb{Z}^{ {M}^w\times 1}$. 	The BS receives  computation tasks uploaded from its associated vehicles. The task arrivals of vehicles follow an arbitrary stochastic process. Let $a_{k,n}^t$ denote the number of the generated tasks of vehicle $n$ in operation slot $t$, and the aggregated computation workload at BS $m$ is given by $A_{k,m}^t=\sum_{n\in \mathcal{N}^t_m}a_{k,n}^t$. Processing all tasks at BSs with limited computing resources may incur prohibitive high queuing delay, and hence a portion of computation tasks can be dispatched to the remote cloud via backbone networks. Let $x_{k,m}^t $ represent the number of dispatched tasks from  BS $m$ in slice $k$, i.e.,
	\begin{equation}\label{equ:constraint_workload_dispatching}
		x_{k,m}^t\in \{0,1,2,..., A_{k,m}^t\},   \forall m \in\mathcal{M}^w.
	\end{equation}
%	\item 

	%Corresponding task arrival rate is $\left(A_k^{max}-A_k^{min}\right)/2$. 
%Hence, $A_m^t-x_m^t$ represents the amount of tasks processed locally. 
%\end{itemize}
The operation decisions impact service delay at each operation slot, which is analyzed in the following subsection. 

%The network planning and operation decisions should be jointly determined to maximize the system performance. In the following two subsections, we present detailed network planning and operation models, respectively.

\subsection{Service Delay Model}
The service delay includes task offloading delay and task processing delay at either the edge or the cloud. For service $k$, the following delay analysis is adopted.

%\begin{itemize}
%	\item 
{Task offloading delay}: The transmission rate of one subcarrier from vehicle $n$ to its associated BS is given by 
%\begin{equation}
	$R_{n}^t=\beta\log_2\left(1+\frac{P_{v}g_{n}^t}{\beta N_o+\beta I}\right),$
%\end{equation}
 where  $P_{v}$, $g_{n}^t$, $N_o$, and $I$ represent vehicle's transmission power, instantaneous channel gain, noise spectrum density, and interference spectrum density, respectively. With the allocated radio spectrum $y_{k,n}^t b_{k,m}^w $, the task offloading delay of vehicle $n$ is given by  %
% \begin{equation}\label{equ:task_offloading_latency}
 $d_{k,n,o}^t=\frac{\xi_k}{y_{k,n}^t b_{k,m}^w  R_{n}^t}, \forall n\in \mathcal{N}^t_m,	$
% \end{equation}
where $\xi_k$ (in bits)  denotes the task data size of service $k$. 

%\item 
{Edge processing delay}: Given the task dispatching decision, $A_{k,m}^t-x_{k,m}^t$ tasks  are processed at BS $m$. Let $Q_{k,m}^t$ (in bits) denote the amount of the backlogged tasks at BS $m$. Taking task computation delay and queuing delay into account,  edge processing delay at BS $m$ is given by
% 	\begin{equation}	
 		$	d_{k,m, e}^t= \frac{\left(Q_{k,m}^t+ {(A_{k,m}^t-x_{k,m}^t+1)\xi_k}/{2} \right) \eta_k}{c_{k,m}^w F_{e}}, \forall m\in \mathcal{M}^w,$
% 			\end{equation}
where $\eta_k$ (in cycles/bit) denotes task computation intensity of service $k$, and $c_{k,m}^w F_{e}$ is the computing capability of BS $m$ with $c_{k,m}^w$ provisioned edge VMs. The task backlog at BS $m$ is updated by 
%	\begin{equation}
$	Q_{k,m}^{t+1}=\left[ Q_{k,m}^t+(A_{k,m}^t-x_{k,m}^t)\xi_k-{c_{k,m}^w F_{e}T_o}/{\eta_k}\right]^+,$
%\end{equation}
where $\left[x\right]^+=\max\left\{x, 0\right\}$.

%\item 
{Cloud processing delay}: For BS $m$, $x_{k,m}^t$ tasks are dispatched via  backbone networks and then processed at the cloud, whose delay is given by %
%	\begin{equation}\label{equ:cloud_delay}
$d_{k,m,c}^t=d^t_{r}+\frac{\xi_k \eta_k}{h_k^w F_{c}},$
%\end{equation}
where $d^t_{r}$ denotes the round trip time in the backbone network. The second term represents the task processing delay in the cloud. Note that the queuing delay at the cloud is negligible as multi-core cloud servers can parallelly process different tasks.
%\end{itemize}

As such, the average delay for each computation task is given by 
\begin{equation}\label{equ:delay_calculation}
	\begin{split}
	&	D^t_k (\mathbf{x}_k^t, \mathbf{y}_k^t)=\sum_{m\in \mathcal{M}^w} \sum_{n\in \mathcal{N}_m^t} \frac{ d_{k,n,o}^t}{\sum_{m\in \mathcal{M}^w} N^t_m}  \\
		&+   \sum_{m\in \mathcal{M}^w} \frac{ d_{k,m,e}^t  \left(A_{k,m}^t-x_{k,m}^t\right) +  d_{k,m,c}^t x_{k,m}^t }{\sum_{m\in \mathcal{M}^w}A_{k,m}^t}. 
	\end{split}
	%\right.\\
	%		&\left.\right),
\end{equation}
 In the above equation, the first term represents the average task offloading delay for each task, and the second term represents the average task processing delay taking workload distribution between the edge and cloud servers into account. By averaging all operation slots, the average service delay is given by
%\begin{equation}
$	\bar{D}_k^w= \frac{1}{T}\sum_{t=1}^{T}D^t_k(\mathbf{x}_k^t, \mathbf{y}_k^t).$
%\end{equation}

\subsection{Network Slicing Cost Model}
The following network slicing cost model is adopted for slicing performance evaluation, including several  components.

%\emph{Slice operation cost} characterizes, i.e., 
%	\begin{equation}
%	\Phi_d^w=q_d \sum_{m\in \mathcal{M}}  o_m^w
%\end{equation}

%\begin{itemize}
%	\item
	  {Slice deployment cost}: The cost is because running network slices at BSs incurs the overhead of resource virtualization, which is given by
%\begin{equation}
%	\begin{split}
	$	\Phi_d^w=q_d \sum_{m\in \mathcal{M}_s}  o_m^w.$
		%&+ q_p  \sum_{k\in \mathcal{K}}\left( h_k^w +\sum_{m\in \mathcal{M}} \left(o_{m}^w  b_{k,m}^w+o_{m}^w c_{k,m}^w \right)  \right).
%	\end{split}
%\end{equation}
Here, $q_d$ denotes the unit cost of deploying network slices at a BS.

%\item 
{Resource provisioning cost}: The cost component characterizes resource provisioning cost of  radio spectrum resources, edge computing resources, and cloud computing resources. For simplicity, we assume the unit costs of a subcarrier, an edge VM instance, and a cloud VM instance are the same, denoted by $q_r>0$. The resource provisioning cost is given by 
%\begin{equation}
%	\begin{split}
			$	\Phi_p^w%&=q_d \sum_{m\in \mathcal{M}}  o_m^w\\&
				= q_r  \sum_{k\in \mathcal{K}}\left( h_k^w +\sum_{m\in \mathcal{M}} \left(o_{m}^w  b_{k,m}^w+o_{m}^w c_{k,m}^w \right)  \right).$
%	\end{split}
%\end{equation}

%	\item
{Slice adjustment cost}: The cost component characterizes the difference between two subsequent planning decisions, % which includes: 1) Slice deployment cost for deploying a new slice at a BS. Launching a new slice instance at a BS requires transferring the VM image to support the instance and booting the VM instance.  Here, $q_n$ represents the unit price of launching a new slice instance; and 2)
i.e., the cost for adjusting the amount of the reserved spectrum and computing resources. For  computing resources,  VM instances can be resized  via advanced virtualization techniques in practical systems, e.g., Kubernetes~\cite{bernstein2014containers}. Here, $q_s$ represents the unit price of adjusting a unit of reserved network resources. Hence, the slice adjustment cost is given by
	\begin{equation}
		\begin{split}
			\Phi_s^w=&	
%				&+\sum_{k\in \mathcal{K}}\sum_{m\in \mathcal{M}}  q_s\left(\left[h_{k}^w-h_{k}^{w-1}\right]^+\right)\\
 q_s  \mathbbm{1}\left\{o_{k, m}^{w-1}=1 \wedge o_{k, m}^{w}=1\right\} \cdot \sum_{k\in \mathcal{K}} \left(  \left[h_{k}^w-h_{k}^{w-1}\right]^+ \right. \\
&\left. +\sum_{m\in \mathcal{M}}  \left(\left[b_{k,m}^w-b_{k,m}^{w-1}\right]^+ + \left[c_{k,m}^w-c_{k,m}^{w-1}\right]^+  \right)  \right),
%&+ q_n \sum_{m\in \mathcal{M}} \left[o_{m}^w-o_{m}^{w-1}\right]^+ 
%			&\right.\\
%			&\left.\right)+
		\end{split}
	\end{equation}
	where $\mathbbm{1}\left\{\cdot \right\}$ is an indicator function and $ \mathbbm{1}\left\{o_{k, m}^{w-1}=1 \wedge o_{k, m}^{w}=1\right\}$ indicates that slice $k$ is deployed in the previous and current planning windows.  %{Note that the spectrum resource resizing is conducted by resource allocation, and the corresponding cost is negligible.}
	
%\item
	
{SLA revenue}: The cost component characterizes the benefit caused by QoS satisfaction, i.e., the achieved service delay of each slice. %As mentioned before,  $\bar{D}_k^w$ represents the average service delay in slice $k$ at planning window $w$, which is measured at the end of each planning window. 
The  piece-wise SLA revenue function is denoted by 
	\begin{equation}\label{equ:revenue_func}
		%\small
		\Omega_k\left(D\right)=
		\begin{cases}
			q_b ,        &\text{if } D<\theta_k^{'}, \\
			q_b \left(\frac{D-\theta_k^{'}}{\theta_k-\theta_k^{'}}\right),        &\text{if } \theta_k^{'} \leq D \leq \theta_k,	\\
			-q_p, &\text{if } D> \theta_k.
		\end{cases}
	\end{equation}
	Here, $q_b>0$  is the highest unit revenue once a slice's SLA is satisfied, and $q_p>0$ is the unit penalty once the slice's SLA is violated. Obviously, $q_p> q_b$ for discouraging slice's SLA violation. In addition, $\theta_k^{'}<\theta_k$ represents the threshold achieving the highest revenue. For simplicity, we set $\theta_k^{'}=\theta_k/2$ in the simulation. The overall SLA revenue of all slices is given by
%	\begin{equation}
		$	\Phi_q^w=\sum_{k\in \mathcal{K}}	\Omega_k \left(\bar{D}_k^w\right).$
%	\end{equation}

%\end{itemize}

Taking all cost components into account, the overall network slicing cost in the entire slice lifecycle (i.e.,  all  planning windows) is given by
%\begin{equation}\label{equ:system_revenue}
%	\begin{split}
		$	\Phi\left( \mathbf{o}^w, \mathbf{B}^w, \mathbf{C}^w, \mathbf{h}^w, \{\mathbf{x}_k^t, \mathbf{y}_k^t\}_{t\in \mathcal{T}, k\in \mathcal{K}}\right)
			=\sum_{w\in \mathcal{W}}\left(\Phi_d^w+\Phi_p^w+\Phi_s^w -\Phi_q^w\right),$
%	\end{split}
%\end{equation}
which is adopted to evaluate network slicing performance. %Note that the SLA revenue component depends on the service delay, which is analyzed as follows.

\section{Problem Formulation}\label{sec: problem_formulation}
The network slicing problem aims to minimize the network slicing cost via determining  network planning decisions at each planning window and network operation decisions at each operation slot for each slice, which is formulated as:

\begin{subequations}\label{Problem 2}
	\begin{align}
		{\mathbf{P}_0:} \,\,\underset{ \begin{subarray}{c} \{\mathbf{o}^w, \mathbf{B}^w,  \mathbf{C}^w, \mathbf{h}^w\}_{ w \in \mathcal{W}} \\ \{\mathbf{x}_k^t, \mathbf{y}_k^t\}_{t\in \mathcal{T}, k \in \mathcal{K}, w \in \mathcal{W}} \end{subarray} }{\text{min}}\,\,
		& \sum_{w\in \mathcal{W}} \Phi\left(\mathbf{o}^w, \mathbf{B}^w, \mathbf{C}^w, \mathbf{h}^w \right) \nonumber \\
		\text{s.t.}\,\,
		&\eqref{equ: slice deployment}, \eqref{equ:constraint_spectrum resource },  \eqref{equ: spectrum capacity}, \eqref{equ: Cloud resource provision}, \eqref{equ:spectrum_constraint}, \text{and }  \eqref{equ:constraint_workload_dispatching}. 
		%		&  0\leq r_{k,m}^w\leq 1,  \forall k\in \mathcal{K}, m\in \mathcal{M}, w\in \mathcal{W} \label{equ:P2constraint_6}
	\end{align}
\end{subequations}

%Due to network dynamics in the large timescale, e.g., time-varying vehicle traffic, addressing the above long-term subproblem requires network traffic information of all the planning windows, which is difficult to be known \emph{a priori}. To solve the subproblem, we design a planning algorithm based on the RL theory, which can make online planning decisions while accommodating spatial-temporally varying traffic dynamics.  

In Problem $\mathbf{P}_0$, the network planning and operation decision making are coupled in two timescales, which should be jointly optimized. To address the challenge, we first decouple the problem into a large-timescale network planning subproblem and multiple small-timescale network operation subproblems.

%\subsection{Problem Decomposition}
\textbf{Subproblem 1}: \emph{Network planning subproblem} is to minimize the network slicing cost across all the planning windows, which is formulated as: 
\begin{subequations}\label{Problem 2}
	\begin{align}
		{\mathbf{P}_1:} \,\,\underset{ \begin{subarray}{c} \{\mathbf{o}^w, \mathbf{B}^w, \\ \mathbf{C}^w, \mathbf{h}^w\}_{ w \in \mathcal{W}}\end{subarray} }{\text{min}}\,\,
		& \sum_{w\in \mathcal{W}} \Phi\left(\mathbf{o}^w, \mathbf{B}^w, \mathbf{C}^w, \mathbf{h}^w\right) \nonumber \\
		\text{s.t.}\,\,
		&\eqref{equ: slice deployment}, \eqref{equ:constraint_spectrum resource },  \eqref{equ: spectrum capacity}, \text{and } \eqref{equ: Cloud resource provision}. 
		%		&  0\leq r_{k,m}^w\leq 1,  \forall k\in \mathcal{K}, m\in \mathcal{M}, w\in \mathcal{W} \label{equ:P2constraint_6}
	\end{align}
\end{subequations}
Addressing the above subproblem requires network traffic information of all  planning windows, which is difficult to be known \emph{a priori}. To solve it, we leverage an  RL method to design a network planning algorithm, which makes online decisions under spatial-temporally varying vehicle traffic.

\textbf{Subproblem 2}: \emph{Network operation subproblem} is  to schedule network resources of each slice to active vehicles with random task arrivals with the objective of minimizing average service delay, which is formulated as:
\begin{subequations}\label{Problem 0}
	\begin{align}
		{\mathbf{P}_2:}	\underset{\mathbf{x}_k^t, \mathbf{y}_k^t}{\text{min}}\,\,
		&   D^t_k(\mathbf{x}_k^t, \mathbf{y}_k^t) \nonumber \\
		\text{s.t.}\,\,
		%	& 	\lim\limits_{T\to \infty}\frac{1}{T}   \sum_{t\in \mathcal{T}} \gamma \eta x_m^t \leq E_{m}, \forall m \in  \mathcal{M}  \label{equ:P0constraint_1}\\
		& 	\eqref{equ:spectrum_constraint} \text{ and } \eqref{equ:constraint_workload_dispatching}. %\\%\sum_{n=1}^{{N}(t)}y_n(t)=1,  \forall t\in \mathcal{T} \label{equ:P0constraint_3}\\
		%		& \sum_{c=1}^{C} z_{n, c}(t) = 1, \forall n\in \mathcal{N}(t), t\in \mathcal{T} \label{equ:P0constraint_5}\\
		%			& x_n(t)\in \mathbb{Z}^+\label{equ:P0constraint_7} \\
		%			& 	0\leq x_n(t) \leq A_n(t),  \forall n\in \mathcal{N}(t), t\in \mathcal{T} \label{equ:P0constraint_2}\\
		%			& 	y_n^t \in \mathbb{R},  \forall n\in \mathcal{N}(t), t\in \mathcal{T} \label{equ:P0constraint_6}\\
		%				&   \label{equ:P0constraint_4}\\
		%					&\
	\end{align}
\end{subequations}

In the above subproblem, radio spectrum resource allocation and task dispatching decisions jointly impact the service delay performance. To solve the problem, we analyze the subproblem property and design an optimization algorithm to make real-time network operation decisions.

%optimizing the long-term service delay performance requires network information in all the operation slots, e.g., vehicles' channel conditions and task arrivals. Hence, it is challenging to determine the optimal network operation decisions. 

%In addition, solving the operation problem via traditional Markov decision process based solution is very challenging, due to (i) prohibitive computational complexity when the dimension of state-action space is large; (ii) time-varying set of associated vehicle users that makes the action and state space varies over time. 

%In the above subproblem, optimizing the long-term service delay performance requires network information in all the operation slots, e.g., vehicles' channel conditions and task arrivals. Hence, it is challenging to determine the optimal network operation decisions. 

%------------------------------------------------------------------------------------------------------------------------------------------------------------------------------------------------------------------------------------------------------------------------------------------------
\section{Learning-Based Network Slicing Algorithm}\label{sec:solution}
In this section, we solve two subproblems in Sections~\ref{subsec: Operation} and~\ref{subsec: planning}, respectively. Finally, we present the TWAS algorithm for jointly optimizing planning and operation decisions in Section~\ref{subsec: twas}.  
%In this section, we propose the learning network slicing algorithm, namely TAWS. TAWS operates in a closed-loop. It adopts an RL algorithm to make the planning decisions. Given the planning decisions, an optimization algorithm is applied to make operation decisions. In addition, the service delay performance obtained  from network operation decisions is incorporated as the reward of RL algorithm, thereby capturing the interaction between network planning and operation stages.

\subsection{Network Operation Optimization}\label{subsec: Operation}
%\begin{itemize}
%	\item Users associate based on distance;
%	\item task dispatching based on estimated latency;
%	\item   spectrum allocation is orthogonal
%\end{itemize}
%In the following, we propose a heuristic network operation algorithm is to minimize the service delay $D^t$ at each time slot. 
We can observe that the radio spectrum allocation decision only impacts offloading delay component, and the task dispatching decision only impacts the computation delay component. Moreover, both decisions are independent in each BS. Hence, the radio spectrum allocation and  task dispatching decisions can be optimized individually at each BS.
%Since spectrum allocation and task dispatching decisions impact  

%To solve the one-shot optimization problem, we further decouple problem $\mathbf{P} 1'$ into a user association subproblem in the outer layer and two subproblems in the inner layer (i.e., a radio spectrum allocation subproblem and a task dispatching subproblem), which are solved respectively. 
%In this subsection, we omit $t$ for notation simplicity. 

\subsubsection{Radio Spectrum Allocation Optimization}% Let  $\mathcal{N}_m^t$ denote the set of active vehicles covered by BS $m$. For an arbitrary vehicle user $j\in \mathcal{N}_m^t$,  the number of the arrived tasks of is denoted by $\hat{\alpha}_j^t$,  and its spectrum efficiency  is denoted by $\hat{R}_j$. The corresponding  radio spectrum allocation decision is denoted by $\hat{y}_j$. 
From \eqref{equ:delay_calculation}, the radio spectrum allocation optimization problem is equivalent to minimizing the task offloading delay at each BS, i.e.,
\begin{subequations}\label{Problem 1_1}
	\begin{align}
		{\mathbf{P}^{r}_m}: \,\,\underset{\mathbf{y}_k^t}{\text{min}}\,\,
		& 	 \sum_{n \in \mathcal{N}_m^t } \frac{\xi_k}{y_{k,n}^t b_{k,m}^w  R_{n}^t} \nonumber \\
		\text{s.t.}\,\,
		& 	 \eqref{equ:spectrum_constraint}.
	\end{align}
\end{subequations}

The objective function can be proved to be convex since its second-order derivative is positive. In addition, the constraint is convex. Hence, problem $\mathbf{P}^{r}_m$ is a convex optimization problem. Using the Karush-Kuhn-Tucker conditions~\cite{boyd2004convex}, the optimal radio spectrum resource allocation decision is  
\begin{equation}\label{equ:optimal_spectrum_allocation}
	({y}_{k,n}^t)^\star=\frac{\sqrt{1/{{R}_n^t}}}{\sum_{i  \in \mathcal{N}_m^t }\sqrt{1 /{{R}_i^t}}}, \forall n  \in \mathcal{N}_m^t. 
\end{equation}
%The radio spectrum allocation of all vehicle users in the network is given by $ (\mathbf{y}^t)^\star =\cup_{m\in \mathcal{M}}\cup_{j\in \mathcal{N}_m^t} \{ (\hat{y}_j^t)^\star\} $, where $ \mathcal{N}^t= \cup_{m\in \mathcal{M}} \mathcal{N}_m^t$. Such closed-form solution is applied to Algorithm~1 (Line~9).

\subsubsection{Task Dispatching Optimization}  
Similarly, from \eqref{equ:delay_calculation},  task dispatching optimization is to minimize the task processing delay, which is formulated as:
\begin{subequations}\label{Problem 1_2}
	\begin{align}
		{\mathbf{P}^w_m:} \,\,\underset{{x_{k,m}^t}}{\text{min}}\,\,
		& d_{k,m,e}^t  \left(A_{k,m}^t-x_{k,m}^t\right) +  d_{k,m,c}^t x_{k,m}^t  \nonumber \\
		\text{s.t.}\,\,
		& 	\eqref{equ:constraint_workload_dispatching}. 
	\end{align}
\end{subequations}
The above objective function can be rewritten as
\begin{equation}\label{equ:PSI_function}
	\begin{split}
	&	\Psi({x_{k,m}^t}) = d_{k,m,e}^t  \left(A_{k,m}^t-x_{k,m}^t\right) +  d_{k,m,c}^t x_{k,m}^t \\ 
		&=	\frac{\nu_1 \xi_k}{2}(x_{k,m}^t)^2+\left(\nu_2^t-\nu_1 \nu_3-\frac{\xi_k A_{k,m} \nu_1 }{2} \right)x_{k,m}^t \\
		&+\nu_1 \nu_3^t A_{k,m}^t.
	\end{split}
\end{equation}
Here, $\nu_1  = \frac{\eta_k}{c_{k,m}^w F_{e}} >0$, $\nu_2^t = d^t_{r}+\frac{\eta_k \xi_k}{h_k^w F_{c}}$, and $\nu_3 =  Q_{k,m}+\frac{A_{k,m}+1}{2}\xi_k$. Since the second-order derivative of the objective function  $\partial^2 \Psi({x_{k,m}^t})/\partial^2 {x_{k,m}^t} =\nu_1^t \xi_k>0$,  the problem is a convex optimization problem~\cite{boyd2004convex}. The optimal task dispatching decision is given by
\begin{equation}\label{equ:optimal_workload_dispatching}
%	\begin{split}
		(x_{k,m}^t)^\star=\frac{ 2\nu_2^t + {\xi_k \nu_1 A_{k,m} }- 2\nu_1 \nu_3^t}{2\nu_1\xi_k}, \forall m \in \mathcal{M}^w .
%	\end{split}	
\end{equation}

\subsection{Network Planing Optimization}\label{subsec: planning}
The network planning problem is a stochastic optimization problem to minimize the  network slicing cost, which can be transformed into a Markov decision process (MDP)~\cite{wu2020dynamic}. The components of the MDP are defined as follows.
%\begin{itemize}
%	\item 

		\emph{1) Action}, which is consistent with planning decisions, including  slice deployment,   radio spectrum and computing resource provisioning at BSs, and cloud computing resource provisioning, i.e., 
%	\begin{equation}
	$	{A}^w=\{\mathbf{o}^w, \mathbf{B}^w, \mathbf{C}^w, \mathbf{h}^w\}.$
%	\end{equation}
	The action dimension is $M_s+2KM+K$. 
	\emph{2) State}, which includes average vehicle traffic density in the current planning window and the planning decisions in the previous window due to the switching cost. The entire area is divided into $J$ disjoint regions, and the average vehicle traffic density of all regions is denoted by $\Lambda^w\in \mathbb{R}^{J\times 1}$. As such, the state is given by
%	\begin{equation}
	$	{S}^w=\{\Lambda^w, \mathbf{o}^{w-1}, \mathbf{B}^{w-1}, \mathbf{C}^{w-1} , \mathbf{h}^{w-1}\}.$
%	\end{equation}
	The state dimension is $2KM+M+K+J$.
%	\item

	 \emph{3) Reward}, which is defined as the inverse of the network slicing cost in the current planning window, i.e.,
%	\begin{equation}
	$	{R}^w \left(S^w, A^w\right) = -\Phi\left(\mathbf{o}^w, \mathbf{B}^w, \mathbf{C}^w, \mathbf{h}^w\right).$
%	\end{equation}
	Note that minimizing the  network slicing cost is equivalent to maximizing the cumulative reward.   
%\end{itemize}

Upon state $S^w$, the learning agent takes action $A^w$, and the corresponding reward $R^w\left(S^w, A^w\right) $ is obtained, along with the state evolves into new state $S^{w+1}$.  With the above setting, our goal is to obtain an optimal planning policy $\pi^\star \in \Pi$ which makes decisions based on the observed state, thereby maximizing the expected long-term cumulative reward. As such, problem $\mathbf{P}_2$ can be formulated as the following MDP:
\begin{subequations}\label{Problem 2}
	\begin{align}
		{\mathbf{P}_2':} \,\,\underset{\pi \in \Pi}{\text{max}}\,\,
		&\mathbb{E}\left[\lim_{W\rightarrow \infty} \sum_{w=1}^{W}(\varphi)^w   R^w \left(S^w, A^w\right) |\pi \right],
		%	\text{s.t.}\,\,
		%	& \eqref{equ:seamless_coverage_constraint}, \eqref{equ: spectrum capacity}, \eqref{equ: computing_capacity}
		%	& 	o_{k,m}^w\in \{0, 1\},  \forall k\in \mathcal{K}, m\in \mathcal{M}, w\in \mathcal{W}\label{equ:P2constraint_1}\\
		%	&  b_{k,m}^w, c_{k,m}^w  \in \mathbb{Z}^+,  \forall k\in \mathcal{K}, m\in \mathcal{M}, w\in \mathcal{W} \label{equ:P2constraint_2}
	\end{align}
\end{subequations}
where $\varphi >0$ is the discount factor. Since vehicle traffic density is continuous, the action-state space can be prohibitively large. To address this issue, an RL algorithm can be adopted.

\begin{algorithm}[t]	\label{algorithm:DARS}
	\small
	\SetAlgoLined
	
	%	\KwResult{Write here the result }
	%	\textbf{Initialization}: Initialize critic and actor networks, and experience replay buffer $\mathcal{B}$;\\
	\SetKwInOut{Input}{Input}
	%	\Input{}
	\SetKwInOut{Output}{Output}
%	\Output{Network planning decisions $\{\mathbf{O}^w, \mathbf{B}^w, \mathbf{C}^w\}_{\forall w\in \mathcal{W}}$\;}
	\For{training episode =$1, 2, ...$}{
%		Initialize neural networks and obtain initial state $S^0$\;
		\For{planning window $w=1, 2, ..., W$}{
			%		{$\rhd$ Experience generation}\\
			Generate planning decisions via the actor network\;% $A^w=\mu(S^w|\theta)+\mathcal{N}(0, \sigma^2)$\;
			\For{each slice \textbf{in parallel} }
			{
					\For{operation slot $t=1, 2,  ..., T$ }
					{
					\For{each BS \textbf{in parallel}}
					{
						Make radio spectrum allocation and task dispatching decisions by \eqref{equ:optimal_spectrum_allocation} and \eqref{equ:optimal_workload_dispatching}\;
						
					}
				Calculate the instantaneous service delay\;
				}
				Measure the average service delay within the planning window\;
			}
	
			Collect vehicle traffic density of all regions, and observe reward $R^w$ and new state $S^{w+1}$\;
			Store transition $\{S^w, A^w, R^w, S^{w+1}\}$ in the experience replay buffer\;
			%			{$\rhd$ Network update}\\
			Sample a random minibatch of transitions from the experience replay buffer\;
			%			Set $y_i=r_i+\gamma Q'(s_{i+1}, \mu'(s_{i+1}|\theta^{\mu'})|\theta^{Q'})$\;
			Update the weights of neural networks\;
%			Update critic and actor target networks\;
			%$L=\frac{1}{N_m}\sum_{i=1}^{N_m}\left(y_i-Q(s_i,a_i|\theta^{Q})\right)^2$\;
%			Update  by  \;

%			\;

		}
	}
	\caption{ TAWS algorithm.}
\end{algorithm}

\subsection{Proposed TAWS Algorithm}\label{subsec: twas}
%With the above optimization-based network operation algorithm and the RL-based network planning algorithm, 

We present the TAWS algorithm to jointly solve the entire network slicing problem $\mathbf{P}_0$, collaboratively integrating RL and optimization methods. The core idea of TAWS is to adopt an RL method for network planning decision making and an optimization method for network operation decision making. The service delay performance is measured at the end of each planning window and then incorporated into the reward in the RL framework, such that the interaction between network planning and operation stages can be captured. The TAWS algorithm is shown in Algorithm~\ref{algorithm:DARS}. %DARS is built within the actor-critic framework, whose architecture is sketched in Fig. \ref{Fig:architecture}.  For critics, two separate critic networks are adopted to evaluate the policy by estimating Q-values. A smaller estimated Q-value among two critic networks is used to update the critic network, which can avoid the Q-value overestimation issue~\cite{fujimoto2018addressing}. Each actor/critic network has an evaluation network and a target network for training stability.

The RL method is based on the deep deterministic policy gradient (DDPG) algorithm~\cite{hausknecht2015deep, lillicrap2015continuous}, which consists of four  neural networks, i.e., actor evaluation network, critic evaluation network, actor target network, and critic target network. %, which are parametrized with weights $\psi $, $\phi$, $\psi ^{'}$, and $\phi^{'}$, respectively. 
In the initialization phase, all  neural networks and the environment are initialized. The procedure of the TAWS is two-step: 1) Network slicing decisions are generated and executed.  The actor network outputs the planning decisions $A^w$,  which is clipped to feasible decision space. The network operation decisions are generated via the optimization method, and the service delay performance is measured at the end of each planning window. The reward $R^w$ can be obtained and the new state can be observed $S^{w+1}$. The transition tuple $\{S^w, A^w, R^w, S^{w+1}\}$ is stored in the experience replay buffer for updating neural networks; and 2) Neural networks are updated.  A mini-batch of transitions are randomly sampled from the experience replay buffer to update the weights of neural networks. Specifically, the critic network is updated by minimizing the loss function, and the actor network is updated via the policy gradient method. Then, actor and critic target networks are updated by slowly copying the weights of evaluation networks.
%			\begin{equation}\label{equ:loss_function}
%	\begin{split}
%		L(\phi)&=\frac{1}{N_m}\sum_{n=1}^{N_m}\left(R^n+\gamma  Q'\left(S^n, \mu'\left(S^n|\psi'\right)  |\phi'\right)\right.\\
%		&\left.-Q\left(S^n,  A^n|\phi \right) \right)^2,
%	\end{split}
%\end{equation}
%where $N_m$ is the minibatch size. 

%			\begin{equation}\label{equ: actor1_update}
%	\begin{split}
%		\nabla_{\psi }J &\approx \frac{1}{N_m}\sum_{n=1}^{N_m}\nabla_{A}{Q\left(S, A|\phi \right)|}_{{A=\mu\left(S^n|\psi \right)}}\cdot \\
%		& \nabla_{\psi }\mu\left(S^n|\psi  \right).
%	\end{split}
%\end{equation}
%Target networks are updated by slowly copying the weights of evaluation networks. %i.e.,
%			\begin{equation}\label{equ:target_update}
%	\begin{split}
%		{\theta'}&=\tau {\theta}+(1-\tau) {\theta'},\\
%		{\phi'}&=\tau{\phi}+(1-\tau){\phi'},
%	\end{split}
%\end{equation}
%where $0<\tau \ll 1$ is the update rate of target networks.

%------------------------------------------------------------------------------------------------------------------------------------------------
\section{Simulation Results}\label{sec: simulation results}

%\textcolor{red}{compress these two paragraph}
%\subsection{Simulation Setup}
We evaluate the performance of the proposed algorithm on real-world vehicle traffic traces in urban vehicular networks. We consider a 1,000$\times$1,000\;m$^2$ simulation area, which is covered by two SBSs and an MBS. Each SBS has a coverage radius of 300\;m, and the MBS located in the centre covers the entire simulation area. %The channel model between BSs and vehicles is given in~\cite{liang2019spectrum}.  
The vehicle traffic density of the simulation area is measured by a unit of a small region of 250$\times$250\;m$^2$, i.e., $J=16$. This dataset is collected by Didi Chuxing GAIA Initiative\footnote{Didi Chuxing Dataset: https://gaia.didichuxing.com.} and contains vehicle traces in the second ring road in Xi’an collected from taxis that are equipped with GPS devices. The periods of a planning window and an operation slot are set to 10\;minutes and 1\;second, respectively. The period of the slice lifecycle is set to 4 hours, including 24 planning windows. The task arrivals of two services both follow Poisson processes with different task arrival rates. We construct two slices for supporting two types of delay-sensitive services. One is an object detect service whose service delay requirement is  100\;\emph{ms}, while the other is  an in-vehicle infotainment service whose service delay requirement is 200\;\emph{ms}. Regarding the TWAS algorithm, the neuron units in hidden layers of both actor and critic networks are set to 128 and 64.  Important simulation parameters are summarized in Table~\ref{Neural network parameters}.

%
% which is collected every 2-4 seconds.\footnote{Note that the data samples at each operation time slot are obtained via an interpolation method.} 

%The vehicle traffic within one hour is plotted in Fig.~\ref{fig:vehicle_traffic}.

% The amount of spectrum resource in a subcarrier is set to 20 MHz, and the CPU frequency of a VM instance is set to 20 GHz. The CPU frequency of the cloud server is set to 200 GHz. The round trip time is set to 0.15\;second. The amount of backhaul transmission rate and energy consumption constraint of each SBS are set to 500\;Mbit/s and 0.2 Joule/slot, respectively, while that of a MBS is set to 800\;Mbit/s, and 0.3 Joule/second, respectively. The task arrivals from vehicles follow a Poisson process with rate $\lambda$. The data size and computation intensity of the task is set to 0.6 Mbit and 1.2$\times 10^8$ CPU cycles. $V$ is set to 5 for balancing service delay performance and energy consumption performance.  For the RL framework, the learning rates of actor and critic networks are set to 5$\times 10^{-4}$ and 5$\times 10^{-3}$, respectively. The neuron units in hidden layers of both actor and critic networks are set to  128 and 64. 
 
%The parameters of the proposed algorithm are given in Table~xxx.

%We compare the proposed scheme with the following benchmark schemes:
%\begin{itemize}
%	\item Edge-only solution
%	
%	\item Myopic optimization solution
%\end{itemize}

\begin{table}[t]
	\tiny
	\centering
	\caption{Simulation Parameters.}
	\label{Neural network parameters}
		\vspace{-0.3cm}
	\begin{tabular}{ c c|c c }
		\hline
		\hline
		\textbf{Parameter} & \textbf{Value} & \textbf{Parameter} & \textbf{Value}\\ \hline
		$N_o$ & $-174$\;dBm& $I$ & $-164$\;dBm \\
		$P_v$ & 27\;dBm &$\beta$& 20\;MHz  \\
		$d_r$ & 0.15 sec & $J$& 16 \\
		%		$F_m$ & \{25, 40\} GHz & $B_m$&\{25, 40\} MHz\\
%		$R_b$ & \{500, 800\} Mbit/s  & $E_m$&\{0.2 0.3\} Joule/slot\\
		$T_o$ & 1 sec &   $T_p$ & 10 min \\
		$F_c$ & 100 GHz  & $F_e$ & 10 GHz  \\
%		$\eta$ & 1.2$\times 10^8$\; CPU cycles   & $\xi $& 0.6 Mbit\\
		$B_m$ & 10 &$C_m$ & 10\\
%		$\gamma$ & 0.24 & $\delta$ & 0.01\\
$\xi_1, \xi_2$& \{0.6, 2\}\;Mbit & $\eta_1, \eta_2$ & \{$1000, 200$\}\;cycles/bit\\
%$\xi_2$& 2\;Mbit & $\eta_2$ & $200$ CPU cycles/bit\\
$\theta_1, \theta_2$ & \{100, 200\}\;ms  & $\theta_1^{'}, \theta_2^{'}$ &  \{50, 100\}\;ms \\\hline
%$$ & 200 ms  & $\theta_2^{'}$ & 100 ms\\
%		$q_r$ & 1 & $q_d$& 10\\
%		$q_s$ & 5& $q_p$& 200\\
%		$q_r$ & 10 &  & \\\hline
	\end{tabular}
		\vspace{-0.2cm}
\end{table}

%\begin{figure}[t]
%	\centering
%	\renewcommand{\figurename}{Fig.}
%	\includegraphics[width=0.3\textwidth]{fig/convergence.eps}
%
%	\caption{Convergence performance of the proposed algorithm.}
%	\label{fig:converegence}
%			\vspace{-0.5cm}
%\end{figure}

\begin{figure}[t]
		\vspace{-0.2cm}
	\centering
	\renewcommand{\figurename}{Fig.}	
	\begin{subfigure}[Convergence]{
			\label{fig:converegence}
			\includegraphics[width=0.22\textwidth]{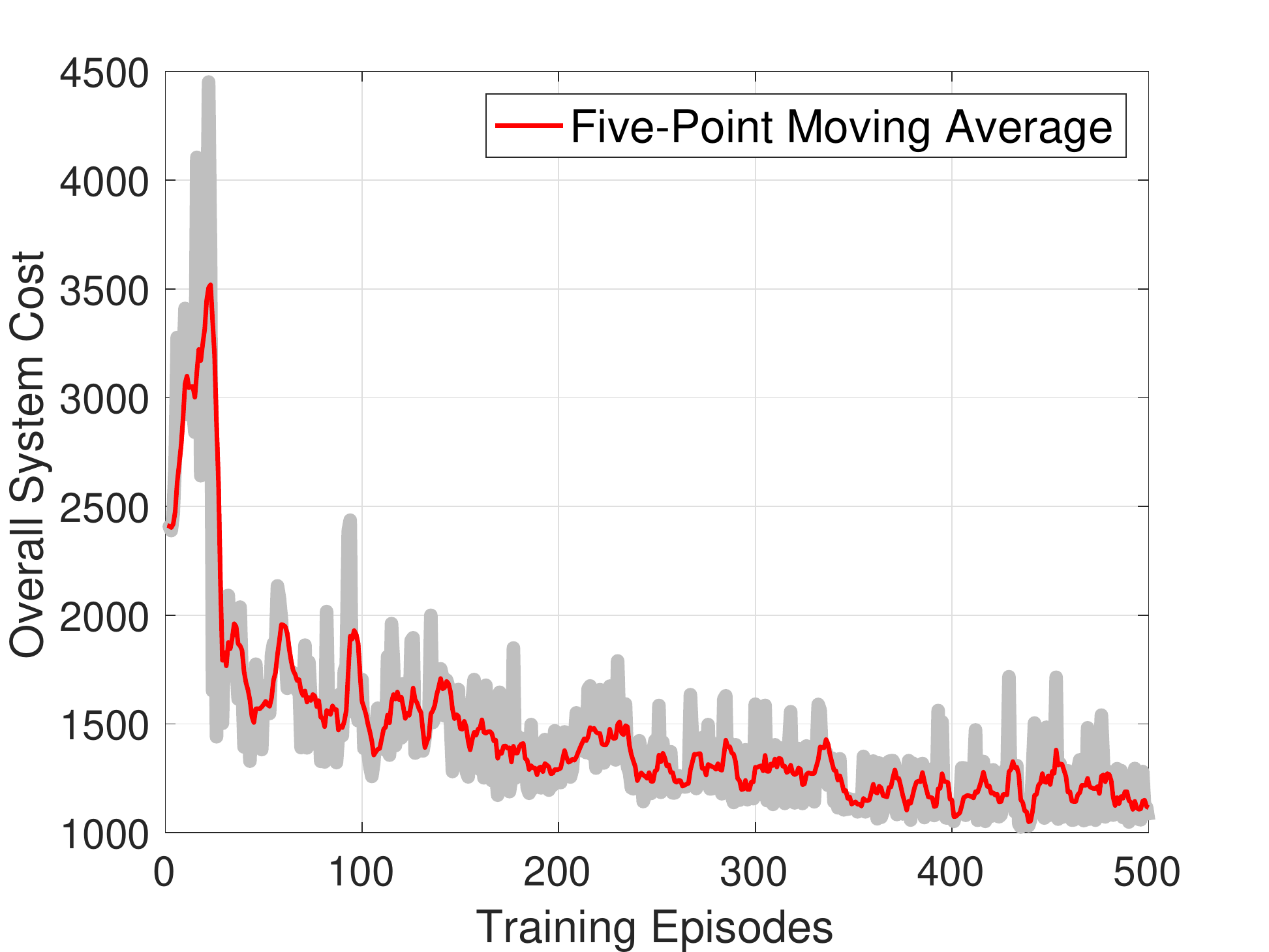}}
	\end{subfigure}
	~
	\begin{subfigure}[Network slicing cost]{
			\label{fig:planning_benchmark}
			\includegraphics[width=0.22\textwidth]{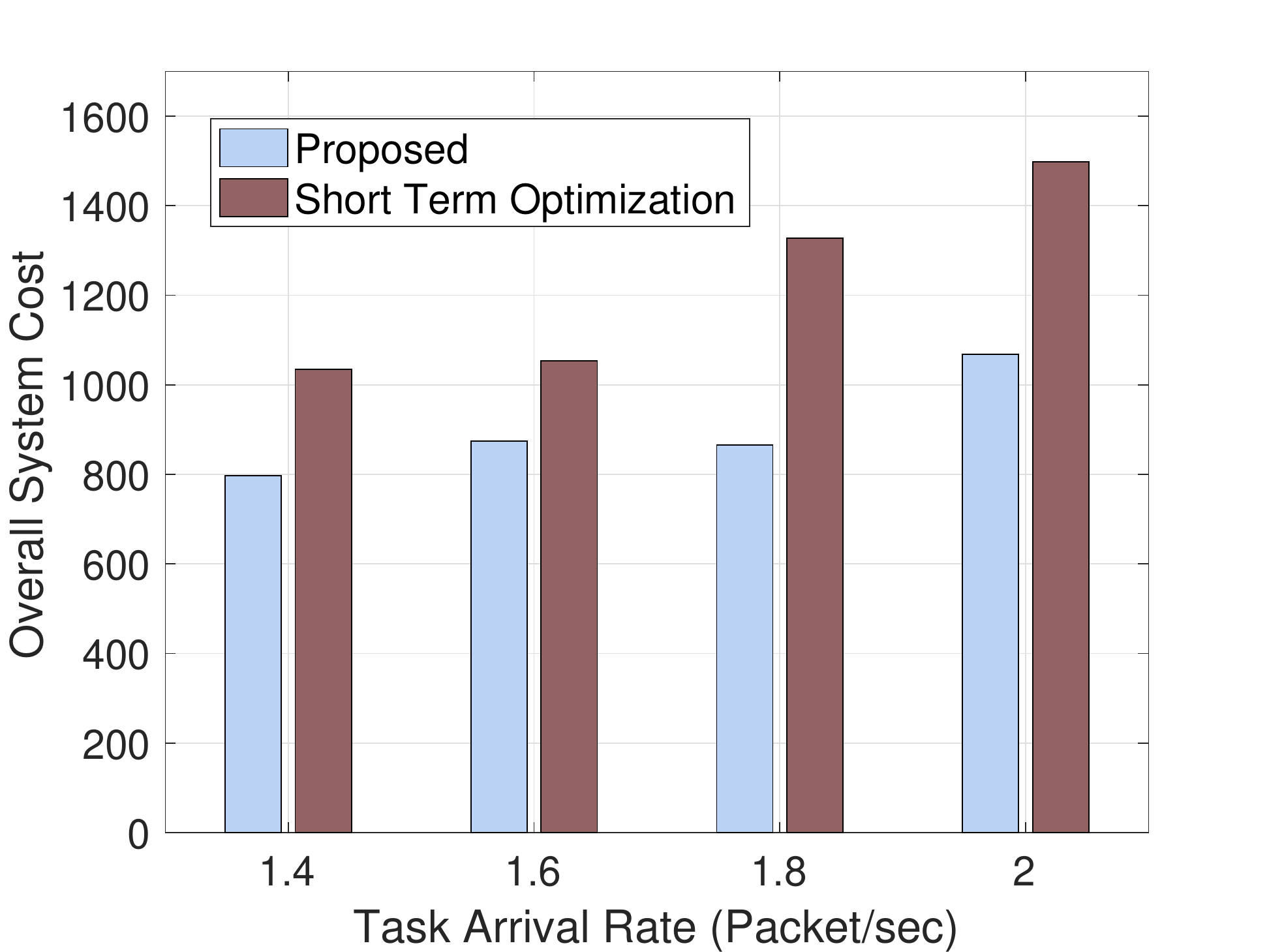}}
	\end{subfigure}
	%		\vspace{-0.3cm}
	\caption{{Performance of the proposed TWAS algorithm.}}
	\label{Fig:Convergence}
		\vspace{-0.5 cm}
\end{figure}

%\emph{Convergence}: 
%\subsection{Performance Evaluation}
As shown in Fig.~\ref{fig:converegence}, we present the overall network slicing cost with respect to training episodes. All simulation points are processed by a five-point moving average in order to highlight the convergence trend of the proposed algorithm. It can be seen that the proposed algorithm converges after 500 training episodes.

% Fig.~\ref{Fig:propvision_cost_converegence} presents the resource provision cost with respect to training episodes, whose results also validate the convergence of the proposed learning solution.

%\begin{figure}[t]
%	\centering
%	\renewcommand{\figurename}{Fig.}
%	\includegraphics[width=0.3\textwidth]{fig/planning_benchmark.eps}
%	%	\vspace{0cm}
%	\caption{Performance comparison with respect to different task arrival rates.}
%	\label{fig:planning_benchmark}
%			\vspace{-0.5cm}
%\end{figure}

As shown in Fig.~\ref{fig:planning_benchmark}, we compare the performance of the proposed algorithm  and a short term optimization benchmark. The basic idea of the benchmark is to minimize the network slicing cost at each individual planning window. Since planning decisions are discrete, a simple exhaustive searching method is adopted to obtain the optimal one-shot planning decisions. Firstly, it can be seen that the proposed algorithm can greatly reduce the network slicing cost as compared to the benchmark. Specifically, when the task arrival rate is 2 packets per second, the proposed algorithm can reduce the  network slicing cost by 23\%. The reason is that the proposed algorithm takes the switching cost between two consequent planning windows into account, while the benchmark scheme does not. Secondly, the overall network slicing cost increases with the increase of the task arrival rate, because more radio and computing resources are consumed in heavy traffic scenarios.

%------------------------------------------------------------------------------------------------------------------------------------------------
\section{Conclusion}\label{sec: conclusion}
In this paper, we have investigated a network slicing problem in edge-cloud orchestrated vehicular networks. A two-stage network slicing algorithm, named TWAS, has been proposed to jointly make network planning and operation decisions in an online fashion. The TAWS can adapt to network dynamics in different timescales, including spatial-temporally varying vehicle traffic density and random task arrivals. Simulation results demonstrat that the TAWS can reduce the network slicing cost as compared to the conventional scheme. For the future work, we aim to determine the optimal planning window size for minimizing the network slicing cost under vehicular network dynamics.

 %The proposed scheme can capture the interaction among decisions in two timescales, which can be applied to other two-timescale resource management problems. 

\bibliographystyle{IEEEtran}
\bibliography{security}

\end{document}